\documentclass[]{spie}  

\usepackage{amsmath,amsfonts,amssymb}
\usepackage{graphicx}
\usepackage{xcolor}
\usepackage[multi-part-units=single]{siunitx}
\usepackage{color}
\usepackage{bm}
\usepackage{url}
\usepackage[export]{adjustbox}
\usepackage{multirow}
\usepackage[colorlinks=true, allcolors=blue]{hyperref}

\title{Unsupervised Anomaly Detection in 3D Brain MRI using Deep Learning with Multi-Task Brain Age Prediction}

\author[a]{Marcel Bengs$^*$}
\author[a]{Finn Behrendt$^*$}
\author[a]{Max-Heinrich Laves}
\author[b]{Julia Krüger}
\author[b]{Roland Opfer}
\author[a]{Alexander Schlaefer}
\affil[a]{Institute of Medical Technology and Intelligent Systems, Hamburg University of Technology, Am
Schwarzenberg-Campus 3, Hamburg 21073, Germany;} 
\affil[b]{jung diagnostics GmbH, Germany}

\authorinfo{Further author information: (Send correspondence to Marcel Bengs)\\ Marcel Bengs: E-mail: marcel.bengs@tuhh.de\\
Alexander Schlaefer: E-mail: schlaefer@tuhh.de\\ $^*$ Both authors contributed equally.}

\begin{document} 
\maketitle 
\fontsize{10}{10}\selectfont
\begin{abstract} 
Lesion detection in brain Magnetic Resonance Images (MRIs) remains a challenging task. MRIs are typically read and interpreted by domain experts, which is a tedious and time-consuming process. Recently, unsupervised anomaly detection (UAD) in brain MRI with deep learning has shown promising results to provide a quick, initial assessment. So far, these methods only rely on the visual appearance of healthy brain anatomy for anomaly detection. Another biomarker for abnormal brain development is the deviation between the brain age and the chronological age, which is unexplored in combination with UAD. We propose deep learning for UAD in 3D brain MRI considering additional age information. We analyze the value of age information during training, as an additional anomaly score, and systematically study several architecture concepts. Based on our analysis, we propose a novel deep learning approach for UAD with multi-task age prediction. We use clinical T1-weighted MRIs of 1735 healthy subjects and the publicly available BraTs 2019 data set for our study. Our novel approach significantly improves UAD performance with an AUC of 92.60\% compared to an AUC-score of 84.37\% using previous approaches without age information.
\end{abstract}


\section{Introduction}
Brain Magnetic Resonance Images (MRIs) are widely used in research and clinical practice for the diagnosis and treatment of neurological diseases.  However, reading and interpreting MRI remains a challenging task. So far, diagnosis requires the manual assessment of the MRIs by trained physicians which is time and resource consuming and especially error-prone in case of unexpected anomalies \cite{bruno2015understanding}. Hence, a fast and accurate decision support tool would have the potential to reduce the workload for physicians, save costs and significantly improve the diagnostic procedure. Recently, supervised deep learning has shown promising results for automatic detection and segmentation of brain lesion in MRI \cite{lundervold2019overview}. However, these methods require large-scale data sets with pixel-level annotations, which are costly and time-consuming to obtain \cite{baur2020autoencoders}. Also, it is typically easy to collect data from healthy subjects compared to data that represents rare diseases, while supervised deep learning is particularly challenging in the case of unbalanced data \cite{fernando2020deep}. \\ In contrast, deep learning for unsupervised anomaly detection (UAD) in brain MRI has been recently proposed, which only relies on healthy brain MRI data \cite{baur2020autoencoders}. Here, the underlying concept for anomaly detection is typically that a network learns to compress and reconstruct the anatomical features of a healthy brain but will not be able to reconstruct unseen anomalies. While a wide range of deep learning methods have been recently proposed for UAD \cite{baur2020autoencoders}, these methods  only learn from image information, i.e., the appearance of healthy brain anatomy in MRI. However, additional patient information such as age is typically considered by radiologists for diagnosis \cite{franke2019ten}, as the deviation between brain age and the actual chronological age can be considered a biomarker for brain abnormalities. However, although chronological age is almost always available, and can be used to characterize deviations from normal brain development, so far it has not been considered for UAD in brain MRI with deep learning. Also, recently it has been demonstrated that brain age may be estimated from MRI with deep learning \cite{jonsson2019brain,levakov2020deep,wang2019gray,zhao2019variational,mouches2021unifying} and yet these findings have not been used for UAD. This opens up the question whether additional age information helps the task of UAD. We systematically study deep learning for UAD in 3D brain MRI combined with age information. We consider age information as an additional input that affects the latent space and as an additional task that allows for brain age analysis. Also, we evaluate a novel combined anomaly score that also considers the difference between predicted brain age and chronological age. 

\section{Methods}

\subsection{Data Sets}
We use a data set with T1-weighted 3D MRIs of 1735 healthy subjects (mean age $45.44 \pm 16.85$ years) from different scanners of 22 vendors and with different field strengths (1T,1.5T, and 3T). The scans were acquired during clinical routine with a standard 3D gradient echo sequence and 68 different scanner configurations. All MRIs have been carefully reviewed by experts and are anonymized. 
 For evaluation of the anomaly detection performance, we consider the publicly available BraTS 2019 data set \cite{menze2014multimodal,bakas2017advancing,bakas2018identifying} and use a subset of 240 MRIs (mean age $60.31 \pm 12.85$ years) from the 335 available MRIs, as age information is not available for the entire data set. We evaluate our methods based on sample level anomaly detection performance, i.e., whether a 3D MRI of a subject is abnormal. Hence, all MRIs, which are ensured to only contain data from healthy subjects are assigned the label healthy/normal and all MRIs of the BraTs 2019 data set are assigned the label unhealthy/anomalous.
We follow the preprocessing steps of a comparative study on different UAD methods \cite{baur2020autoencoders}, including skull stripping, denoising, and standardization. We interpolate all scans to the same isotropic voxel resolution of $1\mathrm{\,mm}\times 1\mathrm{\,mm} \times 1\mathrm{\,mm}$, and crop excessive background by using brain masks of the MRIs. For computational efficiency we downsample our volumes by a factor of 2.5, i.e., from $160 \times 192 \times 166$ to the size of $64 \times 77 \times 66$ voxels. 
\\ We split our data stratified with respect to the subject's age. For training, we use 1255 MRIs of healthy subjects. For validation, we use 80 MRIs of healthy subjects combined with 40 MRIs from the BraTs 2019 data set. For testing, we use 400 MRIs of healthy subjects combined with 200 MRIs from the BraTs 2019 data set.  

\begin{figure}
\centering
\includegraphics[width=0.85\textwidth]{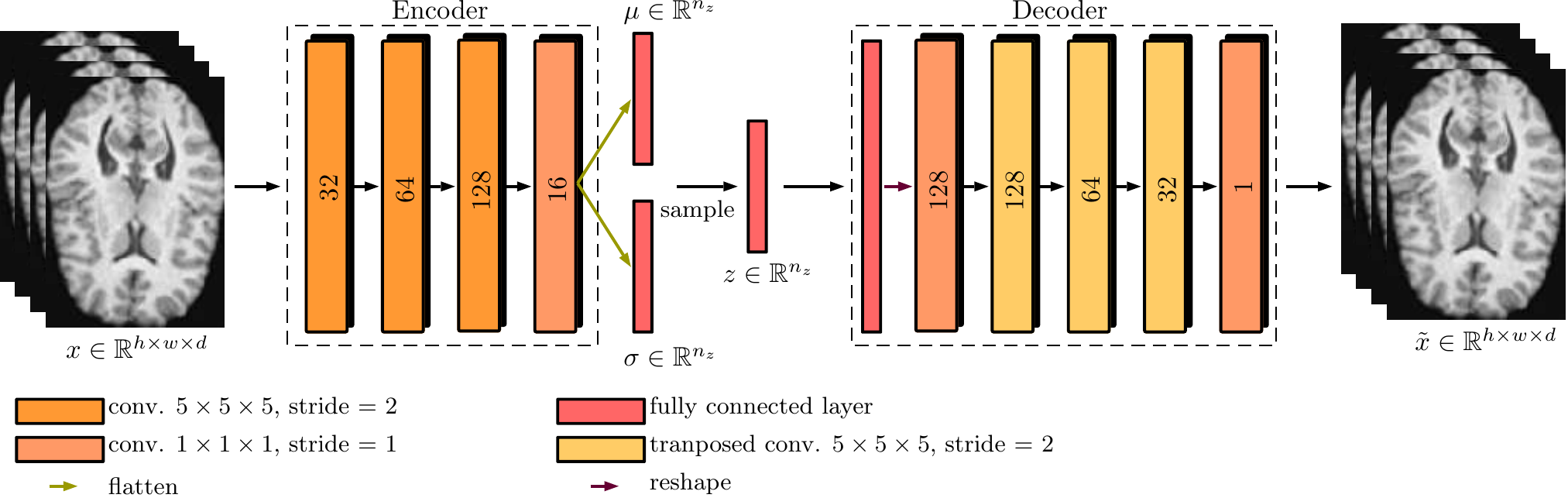}
\includegraphics[width=0.85\textwidth]{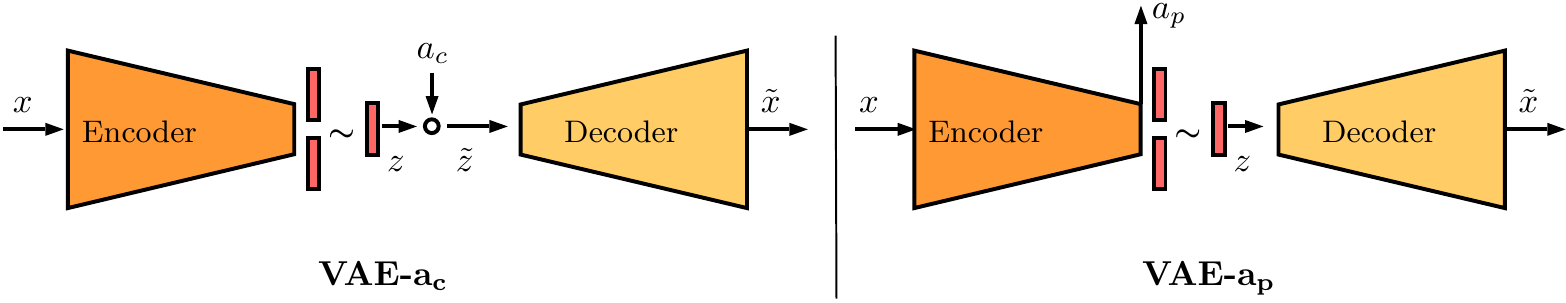}
\caption{Our VAE architecture for UAD and our different architecture concepts for considering additional age information. Note, $a_{c}$ refers to the chronological age and $a_{p}$ refers to the predicted brain age.}
\label{fig:model}
\end{figure}

\subsection{Deep Learning Methods}
As a baseline method we consider the concept of variational autoencoders (VAEs) \cite{kingma2013auto}, which have shown promising results in previous works on UAD in brain MRI \cite{zimmerer2019unsupervised,baur2018deep,baur2020autoencoders}, while being easy to optimize, involving few hyperparameters, and providing fast inference time. The architecture of our VAE is adapted from a recent comparative study on UAD \cite{baur2020autoencoders}. Our VAE and our different methods for considering age information are shown in Figure \ref{fig:model}. We learn directly from entire volumetric T1-weighted MRIs, hence we use 3D networks for all our experiments \cite{bengs2021three}. We study two concepts for combining deep learning for UAD in brain MRI with age information.
First, we consider the chronological age $a_{c}\in\mathbb{R}_{+}$ as an additional network input that affects the latent space representation. To this end, we transform the latent space by multiply the age element-wise with  $z$, hence $\tilde{z}_{i}= z_{i} \cdot a_{c} $. We call these method VAE-$a_{c}$. 
Second, we consider age prediction as an additional task and branch off a regression layer before the latent space of the VAE, as shown in Figure \ref{fig:model}. We refer to this multi-task architecture concept as VAE-$a_{p}$. For all VAEs, we set the latent space size to $n_{z}=2048$, determined based on preliminary experiments using our validation set. \\ Considering the concept of VAEs we  train the baseline without any age information using the following objective.
\begin{equation}
\mathcal{L}=L_{Rec}(\tilde{x},x)+L_{KL}(q(z\vert x),p(z))
\label{eq:VAE_train}
\end{equation}
$L_{KL}$ describes the Kullback–Leibler divergence (KL-Divergence) between the parametrized latent distribution $q(z\vert x)\sim\mathcal{N}(z_{\mu},z_{\sigma})$ and the prior $p(z)$ which follows a multivariate normal distribution \cite{kingma2013auto}. $L_{Rec}$ describes the reconstruction error where we use the voxel-wise $l_{1}$-distance between the input volume $x$ and the output volume, i.e., reconstruction $\tilde{x}$. 
Then, we train our methods with $a_{c}$ as an additional network input, while the objective remains the same as for the baseline VAE. Last, we consider our multi-task method leading to 
\begin{equation}
\mathcal{L}=L_{Rec}(\tilde{x},x)+\beta L_{KL}(q(z\vert x),p(z)) + \gamma L_{Age}(a_{p},a_{c})
\label{eq:train}
\end{equation}
with weighting factors $\beta, \gamma\in\mathbb{R}_{+}$. $L_{Age}$ describes the additional loss-term for the supervised regression task of brain age prediction. Here, we use the $l_{1}$-distance between $a_{p}$ and $a_{c}$. We set $\beta=1$ similar to previous works on UAD with VAEs \cite{zimmerer2019unsupervised,baur2020autoencoders,zimmerer2018context} and also set $\gamma=1$. No hyperparameters are specifically tuned and we train our methods for 400 epochs with standard learning rate of $l_{r}=0.001$ and a batch size of 32 using Adam for optimization. \\  
After training with healthy MRIs only, anomalies in a test image can be detected by considering different anomaly scores. Typically, $L_{Rec}$ or  $L_{KL}$ are used \cite{zimmerer2019unsupervised,baur2020autoencoders}. Our approach VAE-$a_{p}$ also provides an additional anomaly score $L_{Age}$, the deviation between the predicted brain age and the chronological age. We study the discriminative performance of the individual scores and also consider a combined anomaly score.
\begin{equation}
\mathcal{L}_{score}=\alpha_{a} L_{Rec}(\tilde{x},x)+\beta_{a} L_{KL}(q(z|x),p(z)) + \gamma_{a} L_{Age}(a_{p},a_{c})
\end{equation}
Since the different anomaly scores differ by several orders of magnitude we introduce weighting/scaling factors $\alpha_{a}, \beta_{a}, \gamma_{a}\in\mathbb{R}_{+}$ for the combination, and use our validation set performance to utilize a small grid search to find the scaling factors with $\alpha_{a}\in \{0$,$0.001$,\dots,$2\}$, $\beta_{a} =$ $1$, and $\gamma_{a}\in \{0$,$0.01$,\dots,$20\}$. All our methods are implemented in PyTorch (v1.7.0) and trained on a NVIDIA Tesla V100-32GB.

\section{Results}
\begin{table}[t]
\centering
\caption{Comparison of the different methods for anomaly detection. $\mathrm{Specificity}_{(100\%)}$ refers to the best possible specificity for a sensitivity of 100\%. CIs are provided in brackets. $\mathrm{MAE}_{N}$ and $\mathrm{MAE}_{A}$ refer to the MAE between the predicted brain age and the chronological age for healthy subjects and subjects with abnormalities, respectively.} 
\begin{tabular}{lccccccc}
  Method & \bfseries AUC (\%)&  \bfseries AUPRC (\%)  &  \boldsymbol{$\mathrm{Specificity}_{(100\%)}$} \textbf{(\%)} & \boldsymbol{$\mathrm{MAE}_{A}$} \textbf{(years)} &\boldsymbol{$\mathrm{MAE}_{N}$} \textbf{(years)}   \\ 
      \hline
    VAE \cite{zimmerer2019unsupervised,baur2018deep}  &  84.37 (80,87)  & 69.60 (62,76)  &  25.75 (20,31) & N/A & N/A \tabularnewline  
    VAE-$a_{c}$  &    87.65 (85,90) &  76.73 (70,82) & 34.75 (28,44) & N/A & N/A  \tabularnewline
    VAE-$a_{p}$  &  $\pmb{92.60~(90,94)}$ & $\pmb{83.78~(76,88)}$ & $\pmb{43.50~ (37,48)}$ & $14.57\pm 9.72$& $7.13\pm 5.77$  \tabularnewline
 
\label{tab:comparison_results}
  \end{tabular}
\end{table}

\begin{figure}[t]
\centering
\includegraphics[width=0.30\textwidth]{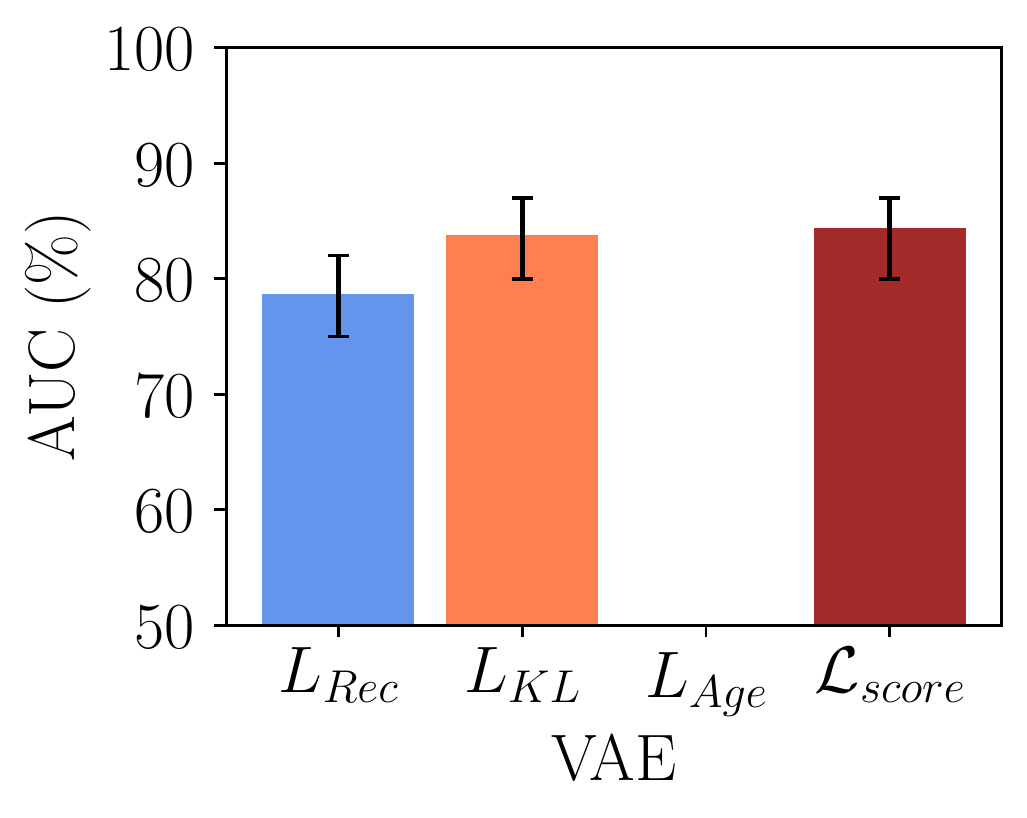}
\includegraphics[width=0.30\textwidth]{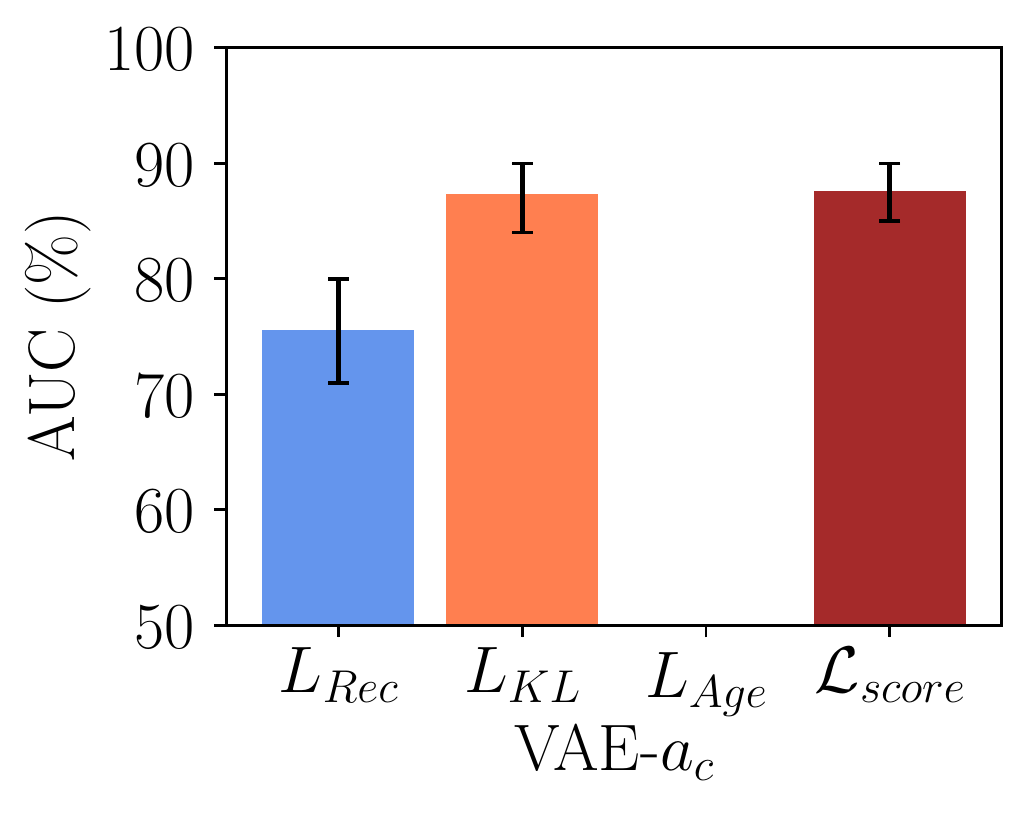}
\includegraphics[width=0.30\textwidth]{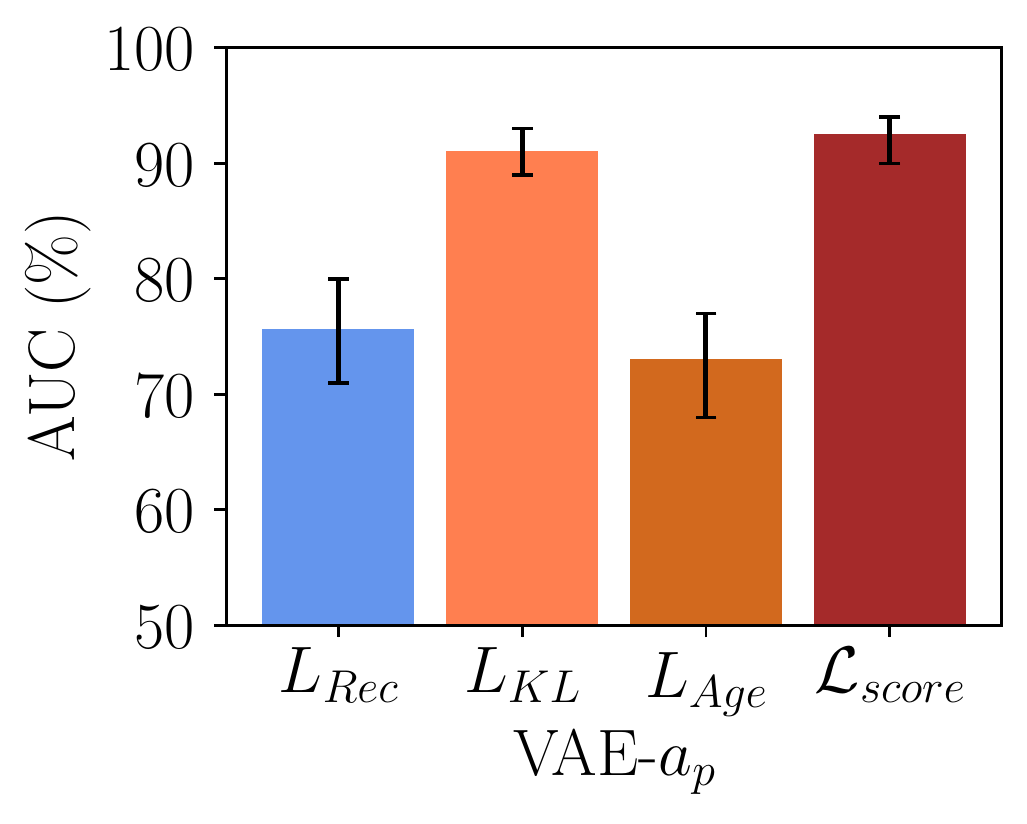}

\caption{Discriminative performance of the anomaly scores. Note, VAE and VAE-$a_{c}$ do not provide age prediction and hence $L_{Age}$ is not available. CIs are shown as error bars.}
\label{fig:bar_chart}
\end{figure}

The anomaly detection performance for the methods is shown in Table  \ref{tab:comparison_results}. For evaluation of our anomaly detection performance independent of the operating point, we consider the area under the ROC curve (AUC) and the area under the precision-recall curve (AUPRC). Considering the clinical application of our methods, we utilize the ROC curves of the models and evaluate the best possible specificity (Specificity$_{(100\%)}$) of the methods for a sensitivity of 100\%. We report $\SI{95}{\percent}$ confidence intervals (CI) using bias corrected and accelerated bootstrapping with $n_{\mathit{CI}} = \num{10000}$ bootstrap samples. For the evaluation of age prediction, we consider the mean absolute error (MAE) for age prediction on normal ($\mathrm{MAE}_{N}$) and abnormal data ($\mathrm{MAE}_{A}$). \\ Compared to our baseline VAE, transforming the latent space with $a_{c}$ leads to around 4\% and 35\% performance improvement considering the AUC and Specificity$_{(100\%)}$, respectively. Considering age prediction as an additional task (VAE-$a_{p}$) leads to 10\% and 67\% performance improvement compared to the baseline VAE considering the AUC and Specificity$_{(100\%)}$, respectively.
Also, our results show that age prediction on normal data can be performed with a $\mathrm{MAE}_{N}=7.13\pm5.77$, while the prediction error is increased by a factor of two on abnormal data. \\ Evaluating the discriminative performance of the anomaly scores in Figure \ref{fig:bar_chart} shows that $L_{KL}$ works best considering the individual anomaly scores, while combining the scores allows for further performance improvements for all methods. 


\section{Discussion and conclusion}
We study UAD in 3D brain MRI using deep learning methods with  brain age information and systematically study the discriminative performance of additional age information. Previously, this task  has  been addressed by considering image information only \cite{baur2020autoencoders,zimmerer2019unsupervised,han2020madgan}. \\ Compared to the baseline VAE, combining the VAE with age information either by transforming the latent space, or by considering age prediction as a multi-task leads to performance improvements in both cases with 35\% and 67\% performance improvements, receptively. It stands out that age prediction as a multi-task works best and is superior to transforming the latent space with $a_{c}$, i.e., providing $a_{c}$ directly as an input. This suggests that the additional task provides effective regularization for UAD. Also, this approach provides an additional anomaly score $L_{Age}$. Our results highlight that the age prediction performance between normal and abnormal data clearly deviates, which further motivates $L_{Age}$ for discrimination. Comparing the discriminative performance of the anomaly scores in Figure \ref{fig:bar_chart} shows that the anomaly score $L_{KL}$ leads to the best individual performance. Combining the scores allows for further performance improvements, similar to previous findings on this topic \cite{zimmerer2019unsupervised}. While $L_{Age}$ allows to improve the performance, the overall performance gains of our methods are not solely based on this additional anomaly score. In fact, using our multi-task concept significantly improves the detection performance of the anomaly score $L_{KL}$. This signifies once again that considering age information as an additional task helps the general concept of UAD. For future work, our approach could be applied to more anomaly detection scenarios and diseases. \\ So far, there is no public benchmark anomaly detection data set available, which would help our work and the field of UAD in general \cite{baur2020autoencoders,zimmerer2019unsupervised}. Considering that our anomalies exclusively constitute brain tumors, our approach and the additional age anomaly score could be even more promising for detecting neurodegenerative and neuropsychiatric diseases, where brain age is more commonly used. Hence, studying our novel approach on a data set with various pathologies that covers the entire age spectrum represents a promising direction for future work. \\ Overall, our results confirm our hypothesis that additional brain age information helps the task of UAD in brain MRI. These results should be considered as age information is simple to obtain and almost always available. \\ \\ Ethical approval: This work was conducted retrospectively on data from clinical routine which was completely anonymized. Ethical approval was therefore not required. Also this work relies on the publicly available BraTs 2019 data set. For use of this data sets, no ethics statements is necessary. \\ \\
\textbf{Acknowledgments.}  This work was partially funded by Grant Number ZF4026303TS9 

\bibliography{report}   

\begin{thebibliography}{10}

\bibitem{bruno2015understanding}
Bruno, M.~A., Walker, E.~A., and Abujudeh, H.~H., ``Understanding and
  confronting our mistakes: the epidemiology of error in radiology and
  strategies for error reduction,'' {\em Radiographics}~{\bf 35}(6),
  1668--1676 (2015).

\bibitem{lundervold2019overview}
Lundervold, A.~S. and Lundervold, A., ``An overview of deep learning in medical
  imaging focusing on mri,'' {\em Zeitschrift f{\"u}r Medizinische Physik}~{\bf
  29}(2),  102--127 (2019).

\bibitem{baur2020autoencoders}
Baur, C., Denner, S., Wiestler, B., Albarqouni, S., and Navab, N.,
  ``Autoencoders for unsupervised anomaly segmentation in brain mr images: A
  comparative study,'' {\em Medical Image Analysis} ,  101952 (2021).

\bibitem{fernando2020deep}
Fernando, T., Gammulle, H., Denman, S., Sridharan, S., and Fookes, C., ``Deep
  learning for medical anomaly detection--a survey,'' {\em arXiv preprint
  arXiv:2012.02364}  (2020).

\bibitem{franke2019ten}
Franke, K. and Gaser, C., ``Ten years of brainage as a neuroimaging biomarker
  of brain aging: what insights have we gained?,'' {\em Frontiers in
  neurology}~{\bf 10},  789 (2019).

\bibitem{jonsson2019brain}
J{\'o}nsson, B.~A., Bjornsdottir, G., Thorgeirsson, T., Ellingsen, L.~M.,
  Walters, G.~B., Gudbjartsson, D., Stefansson, H., Stefansson, K., and
  Ulfarsson, M., ``Brain age prediction using deep learning uncovers associated
  sequence variants,'' {\em Nature communications}~{\bf 10}(1),  1--10 (2019).

\bibitem{levakov2020deep}
Levakov, G., Rosenthal, G., Shelef, I., Raviv, T.~R., and Avidan, G., ``From a
  deep learning model back to the brain—identifying regional predictors and
  their relation to aging,'' {\em Human brain mapping}~{\bf 41}(12),
  3235--3252 (2020).

\bibitem{wang2019gray}
Wang, J., Knol, M.~J., Tiulpin, A., Dubost, F., de~Bruijne, M., Vernooij,
  M.~W., Adams, H.~H., Ikram, M.~A., Niessen, W.~J., and Roshchupkin, G.~V.,
  ``Gray matter age prediction as a biomarker for risk of dementia,'' {\em
  Proceedings of the National Academy of Sciences}~{\bf 116}(42),  21213--21218
  (2019).

\bibitem{zhao2019variational}
Zhao, Q., Adeli, E., Honnorat, N., Leng, T., and Pohl, K.~M., ``Variational
  autoencoder for regression: Application to brain aging analysis,'' in [{\em
  International Conference on Medical Image Computing and Computer-Assisted
  Intervention}{\nolinebreak\hspace{0.1em}]},   823--831, Springer (2019).

\bibitem{mouches2021unifying}
Mouches, P., Wilms, M., Rajashekar, D., Langner, S., and Forkert, N.,
  ``Unifying brain age prediction and age-conditioned template generation with
  a deterministic autoencoder,'' in [{\em Medical Imaging with Deep
  Learning}{\nolinebreak\hspace{0.1em}]},  (2021).

\bibitem{menze2014multimodal}
Menze, B.~H., Jakab, A., Bauer, S., Kalpathy-Cramer, J., Farahani, K., Kirby,
  J., Burren, Y., Porz, N., Slotboom, J., Wiest, R., et~al., ``The multimodal
  brain tumor image segmentation benchmark (brats),'' {\em IEEE transactions on
  medical imaging}~{\bf 34}(10),  1993--2024 (2014).

\bibitem{bakas2017advancing}
Bakas, S., Akbari, H., Sotiras, A., Bilello, M., Rozycki, M., Kirby, J.~S.,
  Freymann, J.~B., Farahani, K., and Davatzikos, C., ``Advancing the cancer
  genome atlas glioma mri collections with expert segmentation labels and
  radiomic features,'' {\em Scientific data}~{\bf 4},  170117 (2017).

\bibitem{bakas2018identifying}
Bakas, S., Reyes, M., Jakab, A., Bauer, S., Rempfler, M., Crimi, A., Shinohara,
  R.~T., Berger, C., Ha, S.~M., Rozycki, M., et~al., ``Identifying the best
  machine learning algorithms for brain tumor segmentation, progression
  assessment, and overall survival prediction in the brats challenge,'' {\em
  arXiv preprint arXiv:1811.02629}  (2018).

\bibitem{kingma2013auto}
Kingma, D.~P. and Welling, M., ``Auto-encoding variational bayes,'' {\em ICLR}
  (2014).

\bibitem{zimmerer2019unsupervised}
Zimmerer, D., Isensee, F., Petersen, J., Kohl, S., and Maier-Hein, K.,
  ``Unsupervised anomaly localization using variational auto-encoders,'' in
  [{\em MICCAI}{\nolinebreak\hspace{0.1em}]},   289--297, Springer (2019).

\bibitem{baur2018deep}
Baur, C., Wiestler, B., Albarqouni, S., and Navab, N., ``Deep autoencoding
  models for unsupervised anomaly segmentation in brain mr images,'' in [{\em
  International MICCAI Brainlesion Workshop}{\nolinebreak\hspace{0.1em}]},
  161--169, Springer (2018).

\bibitem{bengs2021three}
Bengs, M., Behrendt, F., Kr{\"u}ger, J., Opfer, R., and Schlaefer, A.,
  ``Three-dimensional deep learning with spatial erasing for unsupervised
  anomaly segmentation in brain mri,'' {\em International journal of computer
  assisted radiology and surgery}~{\bf 16}(9),  1413--1423 (2021).

\bibitem{zimmerer2018context}
Zimmerer, D., Kohl, S., Petersen, J., Isensee, F., and Maier-Hein, K.,
  ``Context-encoding variational autoencoder for unsupervised anomaly
  detection,'' {\em MIDL}  (2019).

\bibitem{han2020madgan}
Han, C., Rundo, L., Murao, K., Noguchi, T., Shimahara, Y., Milacski, Z.~A.,
  Koshino, S., Sala, E., Nakayama, H., and Satoh, S., ``Madgan: unsupervised
  medical anomaly detection gan using multiple adjacent brain mri slice
  reconstruction,'' {\em arXiv preprint arXiv:2007.13559}  (2020).

\end{thebibliography}
\bibliographystyle{spiebib}   

\end{document}